\documentclass[12pt]{iopart}

\usepackage{iopams}  

\usepackage[superscript,biblabel]{cite}
\usepackage[colorlinks]{hyperref}\hypersetup{
     colorlinks   = true,
     citecolor    = blue
}
\usepackage{hyperref}
\hypersetup{
    colorlinks=true,
    linkcolor=blue,
    filecolor=magenta,      
    urlcolor=cyan,
}
\usepackage[normalem]{ulem}
\usepackage[toc,page]{appendix}
\usepackage[utf8]{inputenc}
\usepackage[T1]{fontenc}
\usepackage{lmodern}
\usepackage[]{babel}
\usepackage{color}
\usepackage{graphics}
\usepackage{epsfig}
\usepackage{float}
\usepackage{array}
\usepackage{bm}
\newcolumntype{C}[1]{>{\centering\arraybackslash$}p{#1}<{$}}
\begin{document}

\title[Kinetic simulations of collision-less plasma in open magnetic geometries]{Kinetic simulations of collision-less plasmas in open magnetic geometries$\dagger$}

\author{Atul Kumar$^1$, Juan F Caneses Marin$^2$}

\address{Oak Ridge National Laboratory, 1 Bethel Valley Road, Oak Ridge, TN-37831, USA}
\ead{$^1$kumara1@ornl.gov, and $^2$canesesmarjf@ornl.gov}
\vspace{10pt}

\begin{abstract}
Laboratory plasmas in open magnetic geometries can be found in many different applications such as (1) Scrape-Of-Layer (SOL) and divertor regions in toroidal confinement fusion devices ($\approx1-10^2\hspace{1mm}\mathrm{eV}$), (2) linear divertor simulators ($\approx1-10\hspace{1mm}\mathrm{eV}$), (3) plasma-based thrusters ($\approx10\hspace{1mm}\mathrm{eV}$) and (4) magnetic mirrors ($\approx10^2-10^3\hspace{1mm}\mathrm{eV}$).  A common feature of these plasma systems is the need to resolve, in addition to velocity space, at least one physical dimension (e.g. along flux lines) to capture the relevant physics. In general, this requires a kinetic treatment. Fully kinetic Particle-In-Cell (PIC) simulations can be applied but at the expense of large computational effort. A common way to resolve this is to use a hybrid approach: kinetic ions and fluid electrons.
In the present work, the development of a hybrid PIC computational tool suitable for open magnetic geometries is described which includes (1) the effect of non-uniform magnetic fields, (2) finite fully-absorbing boundaries for the particles and (3) volumetric particle sources. Analytical expressions for the momentum transport in the paraxial limit are presented with their underlying assumptions and are used to validate the results from the PIC simulations. A general method is described to construct discrete particle distribution functions in state of mirror-equilibrium. This method is used to obtain the initial state for the PIC simulation. 
Collisionless simulations in a mirror geometry are performed. Results show that the effect of magnetic compression is correctly described and momentum is conserved. The self-consistent electric field is calculated and is shown to modify the ion velocity distribution function in manner consistent with analytic theory. Based on this analysis, the ion distribution function is understood in terms of a loss-cone distribution and an isotropic Maxwell-Boltzmann distribution driven by a volumetric plasma source. Finally, inclusion of a Monte-Carlo based Fokker-Planck collision operator is discussed in the context of future work.
\footnotetext[0]{This manuscript has been authored by UT-Battelle, LLC, under contract DE-AC05-00OR22725 with the US Department of Energy (DOE). The US government retains and the publisher, by accepting the article for publication, acknowledges that the US government retains a nonexclusive, paid-up, irrevocable, worldwide license to publish or reproduce the published form of this manuscript, or allow others to do so, for US government purposes. DOE will provide public access to these results of federally sponsored research in accordance with the DOE Public Access Plan (http://energy.gov/downloads/doe-public-access-plan).}\end{abstract}
%
%
%
%
%

\section{Introduction: Context and objectives}
\label{Introduction}
The main objective of the work herein presented is the development of computational tools to model the evolution of the distribution function of plasmas  in open magnetic geometries  in the presence of non-uniform magnetic fields, Radio-Frequency (RF) heating and/or volumetric plasma sources. Important examples of such plasma systems include (1) Scrape-Of-Layer (SOL) and divertor regions in toroidal confinement fusion devices ($\approx1-100\hspace{1mm}\mathrm{eV}$), (2) linear divertor simulators ($\approx1-10\hspace{1mm}\mathrm{eV}$) such as the MPEX\cite{caneses_effect_2020,rapp_materials_2020} , (3) plasma-based thrusters ($\approx10\hspace{1mm}\mathrm{eV}$) such as the VASIMR\cite{bering_electromagnetic_2008} and (4) magnetic mirrors ($\approx10^2-10^3\hspace{1mm}\mathrm{eV}$)  such as the WHAM \cite{Tecdoc2019, Anderson_FST_2021} . 

Understanding, predicting and controlling the velocity-space distribution of these plasma systems and associated parallel transport is required to achieve their scientific/technical objectives.  For example, calculating the SOL heat fluxes and ionization/excitation rates in the divertor region require the inclusion of kinetic effects to correctly describe the contribution of non-thermal distribution functions. Another example is the application of RF heating in linear divertor simulators to increase plasma temperature. In the MPEX concept\cite{caneses_effect_2020,rapp_materials_2020}, this is done via resonant RF heating schemes where the interplay between RF-driven velocity space diffusion, collisional relaxation and magnetic mirror effects determine the evolution of the distribution function.  An important feature of these plasma systems is that in addition to resolving the physics in velocity space, the dimension along the magnetic field needs to be retained. The kinetic description of these systems is best done with the Boltzmann transport equation with suitable source and sink terms. In the context of plasmas, the acceleration term in the Boltzmann transport equation is replaced by the Lorentz force which leads to the so-called Vlasov-Maxwell system of equations.

An important method for solving the Vlasov-Maxwell system of equations is the Particle-In-Cell (PIC) technique which describes both ions and electrons kinetically. Moreover, when ions are treated kinetically while electrons are modeled as a fluid, the technique is refereed to as hybrid PIC. This method allows ions to develop non-thermal distributions and provide means to intrinsically capture many effects that are difficult, if not impossible, to capture in a fluid-ion formulation. Examples of phenomena that may require a non-Maxwellian treatment of ions include: flow interpenetration \cite{WinskeLeroy1984,MANKOFSKY198789,Akimoto1991,Winske1992}, ion-viscosity, non-isotropic pressure \cite{Baranger2011}, acceleration of low density ion beams, finite ion gyroradii effects \cite{Gingell_2012}, diffusion and kinetic mix at shocked interfaces  \cite{Weidl2016,Clark2013,Carbajal2014}, and plasma heating\cite{Gingell_2014,Gingell_2013,Gingell_2012} . Moreover, such hybrid codes are not restricted to resolving electron spatial scales such as the Debye length, the electron skin depth or the electron gyro-radius, and their associated time scales. Thus, it is possible to use these hybrid codes to model much larger spatial/temporal scales and use higher dimensionality than is practical using a fully kinetic method. However, with fluid electrons, inertial electron dynamics or other electron kinetic effects are not resolved. An additional advantage of hybrid over fully-kinetic methods is the presence of a well-defined electron temperature, and thus it is easier to interface the system with physics modules such as equation-of-state (EOS), ionization, and radiation transport. Different flavors of kinetic-ion hybrid particle codes are being used for many years to capture the details of physics related to ion heating \cite{Chodura_1975,Winske1985} magnetized pinches \cite{sagro1976,Hamasaki1977,HARNED1982452}, space plasmas \cite{Loroy1981,DALDORFF2014236,Franci_2015,Vasquez_2012} magnetic fusion energy \cite{MATTHEWS1994102,Todo1998,Liu2001} and high-energy-density physics (HEDP) \cite{Casanova1991,Thoma2017,Keenan2018} . Furthermore, there are many vibrant areas of research in magnetic fusion that relate directly to kinetic-ion physics.  

In this paper, the development of a parallel, electromagnetic, Hybrid Particle-In-Cell (PIC) code is described and employed to simulate plasmas in open magnetic geometries. The code resolves 1 dimension in physical space (1D) and three dimensions in velocity space (3D). In the present work, the collision-less evolution of the ion Probability Density Function (PDF) in a magnetic mirror is investigated as this forms the basis of many plasma systems with non-uniform magnetic field geometries where the physical dimension along the magnetic flux must be retained. Addition of Coulomb collisions, RF heating and volumetric particle/energy sources and sinks to this modelling capability allows simulation of various systems of interest such as those mentioned previously.

This paper is structures as follows: (1) an overview of the hybrid PIC technique is presented followed by (2) a description of important physics required to apply the hybrid PIC to open magnetic geometries. Next, (3) the particle and momentum transport in the paraxial limit are introduced and associated assumptions presented. This is followed by (4) a discussion on setting the initial conditions for the PIC calculations, (5) the presentation of the results and (6) discussion.

\section{Overview of the Hybrid Particle-In -Cell technique} 
\label{HPIC}
The PIC simulation technique is commonly used in plasma physics to study kinetic effects in plasmas not captured in fluid models\cite{BirdsallLangdon,Hockney_1988}; moreover, the use of the \textit{hybrid} PIC approach enables simulation of plasma systems over timescales much longer than practically realizable with fully-kinetic (PIC) simulations at the price of not incorporating a full kinetic treatment of electron dynamics. This level of description is well adapted to phenomena occurring on length scales comparable to, or greater than, the ion inertial length, and on timescales comparable to the ion gyro-period \cite{Clark2013, Winske1996}. The hybrid PIC model describes the different ion populations as a collection of super-particles and the electrons as a neutralizing, mass-less fluid. The dynamics of the ion is governed by the Lorentz force Eq. \ref{ionmotion} and Eq. \ref{ionmom}, where $\mathbf{r_\alpha}$ and $\mathbf{v_\alpha}$ are the position and velocity vector of an ion of species $\alpha$ respectively, and $m_\alpha$ and $Z_\alpha$ are the corresponding mass and charge state. On the other hand, electrons are considered as an isotropic and isothermal ideal gas, $P_e=nT_e$.

\begin{equation}
\frac{d\mathbf{r_\alpha}}{dt} = \mathbf{v_\alpha}
\label{ionmotion}
\end{equation}\begin{equation}
m_\alpha\frac{d\mathbf{v_\alpha}}{dt} =eZ_\alpha[\mathbf{E(r_\alpha)} +\mathbf{v_\alpha} \times \mathbf{B(r_\alpha)}]
\label{ionmom}
\end{equation}
To advance ions in time according to equations(\ref{ionmotion}-\ref{ionmom}), a time-centered. finite difference leapfrog method \cite{Hockney_1988, Lipatov} is employed. The electric $\mathbf{E}$ and magnetic $\mathbf{B}$ fields are linked together by the generalized Ohm's law (Eq. \ref{GOHM}), where $n$ is the plasma density, $\mathbf{J_p}$ the plasma current, $\mathbf{B}$ is the total magnetic field, $\mathbf{U_i}$ the total ion bulk velocity and $P_e$ the electron pressure.
\begin{equation}
\mathbf{E} = \frac{1}{en} \mathbf{J_p} \times \mathbf{B} - \mathbf{U_i} \times \mathbf{B} - \frac{1}{en} \nabla P_e
\label{GOHM}
\end{equation}In the above expression, the total magnetic field $\mathbf{B}$ is composed of the externally applied magnetic field $\mathbf{B}_{0}$ and the plasma contribution $\mathbf{B}_p$ (Eq. \ref{Btotal}). Moreover, the plasma current $\mathbf{J_p}$ is calculated via Ampere's law in the low frequency limit (Eq. \ref{Ampere0}) where $\mathbf{B_p}$ is advanced via Faraday's law (Eq. \ref{Faraday}).
\begin{equation}
\mathbf{B}=\mathbf{B}_0 +\mathbf{B}_p
\label{Btotal}
\end{equation}\begin{equation} 
\nabla \times \mathbf{B_p} = \mu_0\mathbf{J_p}
\label{Ampere0}
\end{equation}\begin{equation}
\frac{\partial \mathbf{B_p}}{\partial t} = -\nabla \times \mathbf{E}
\label{Faraday}
\end{equation} The electric field in Eq.(\ref{GOHM}) is a state quantity \cite{Colonna} which is a function of ions moments: density $n$, ion bulk velocity $\mathbf{U_i}$, magnetic field $\mathbf{B}$ and electron temperature $T_e$. This implies that the electromagnetic fields in the Hybrid PIC approach are solved without specifying field boundary conditions. Moreover, the electromagnetic fields are described using a 1-D staggered grid\cite{KaneYee1966,Buneman1993} which effectively produces a central difference operator with an error of order $\mathcal{O}(\Delta{x}^2)$.   The total ion bulk velocity $\mathbf{U_i}$ is given by Eq. \ref{bulkVel} where the summation is over $k$ species of ions, $\mathbf{u_\alpha}$ and $n_\alpha$ are the bulk velocity and ion number density of species $\alpha$ respectively. 
\begin{equation}
\mathbf{U}_i =\frac{ \sum_{\alpha=1}^{k} (Z_\alpha n_\alpha \mathbf{u_j})}{ \sum_{\alpha=1}^{k} (Z_\alpha n_\alpha)}
\label{bulkVel}
\end{equation}The plasma density $n$ is obtained from the quasi-neutrality condition:
 \begin{equation}
n \equiv n_e = \sum_{\alpha=1}^{k} (Z_\alpha n_\alpha )
\label{den}
\end{equation}
A hybrid PIC model for the multi-species case can be completely defined by Equations (\ref{ionmotion})–(\ref{den}). This model exhibits the existence of low-frequency kinetic electrostatic modes such as ion-Bernstein waves, as well as electromagnetic modes including ion-cyclotron waves and Alfv\'en waves.

It is important to note that solving Eq. \ref{ionmotion} and Eq. \ref{ionmom} in a PIC framework is equivalent to solving the Vlasov equation using the single particle (Klimontovich) Probability Distribution Function (PDF) for each species $\alpha$ presented in Eq. \ref{klimontovich}. In this expression, $N_{SP}^\alpha$ is the total number of super-particles, $N_C^\alpha$ the total number of computational particles and $a_i^\alpha$ is the weight of the $i^{th}$ computational particle of species $\alpha$. Each delta function product represent a particle in phase-space whose trajectory $(\mathbf{r_i^\alpha(t)},\mathbf{v_i
^\alpha (t)})$ is defined by Eq. \ref{ionmotion} and Eq. \ref{ionmom}.
\begin{equation}
f_\alpha (\mathbf{r},\mathbf{v},t)= \frac{1}{N_{SP}^\alpha}\sum_{i=1}^{N_C^\alpha}{a_i^\alpha\delta(\mathbf{r}-\mathbf{r_i^\alpha(t)})\delta(\mathbf{v}-\mathbf{v_i^\alpha(t)})}\hspace{1cm} \mathrm{[m^{-6}s^{-3}]}
\label{klimontovich}
\end{equation}In the present work, the  electromagnetic parallel hybrid PIC code PROMETHEUS++ \cite{Carbajal2014} is employed. PROMETHEUS++ solves for the plasma dynamics in one dimension in physical space and 3 dimensions in velocity space (1D-3V). PROMETHEUS++ is used as the starting point for developing a hybrid modelling capability for plasma systems confined in open magnetic field geometries such as SOL plasmas, linear divertor simulators, electric thrusters and/or mirror devices.

\section{Hybrid PIC for open magnetic geometries}
The 1D-3V hybrid PIC code PROMETHEUS++ as described in reference \cite{Carbajal2014}  assumes uniform profiles along the spatial dimension and periodic particle boundary conditions. For the present study, PROMETHEUS++ has been used as the basis and significantly modified to make it suitable for simulating plasmas in open magnetic geometries. The main modifications include: (1) the effect of externally applied \textit{non-uniform} magnetic fields and (2) presence of finite boundaries and (3) volumetric particle sources. The presence of non-uniform magnetic fields, introduces radial variation in the plasma cross section which effectively produces a 1.5D-3V PIC code. As a result, the code is hereafter referred to as "PRO++ OMG" where OMG is an abbreviation for {\bf O}pen {\bf M}agnetic {\bf G}eometries.

\subsection{Non-uniform magnetic fields}
As described in section \ref{HPIC}, the PIC formulation is equivalent to solving the Vlasov equation using a Klimontovich PDF (Eq. \ref{klimontovich}), where each delta function represents a super-particle in phase-space whose trajectory is governed by the equations of motion shown in Eq. \ref{ionmotion} and Eq. \ref{ionmom}. In a 1D-3V formulation, two out of the three dimensions in physical space are integrated out leading to the reduced Klimontovich PDF shown in Eq. \ref{klimontovichCompressed}, where $A(x_i)$ represents the cross sectional area of the plasma at the position $x_i^\alpha$ and $a_i$ is the weight of each computational particle.

\begin{equation}
f_\alpha (x,\mathbf{v},t)= \frac{1}{N_{SP}^\alpha} \sum_{i=1}^{N_C^\alpha}{\frac{a_i^\alpha}{A(x_i^\alpha)} \delta(x-x_i^\alpha)\delta(\mathbf{v}-\mathbf{v_i^\alpha})}\hspace{1cm} \mathrm{[m^{-6}s^{-3}]}
\label{klimontovichCompressed}
\end{equation}
To illustrate the effect of magnetic compression due to a non-uniform magnetic field, the above PDF is integrated over all velocity space to obtain the zeroth moment (Eq. \ref{zerothMom}) and consequently the particle density $n_\alpha$ for the $\alpha$ species. In Eq. \ref{zerothMom}, the term $N_R^\alpha $ is the total number of real particles of species $\alpha$ represented in the computational domain. Using conservation of magnetic flux in the form $B_{x0} A_0=B_x(x)A(x)$ where $B_{x0}$  and $A_0$ are the magnetic field and plasma cross sectional area respectively at some reference location, the particle density for the $\alpha$ species is given by Eq. \ref{ne_compressed}. The term $B_x(x_i^\alpha)/B_{x0}$ represents the magnetic compression effect caused by the non-uniform externally applied magnetic field. In the case of a uniform magnetic field, the compression term becomes unity and the usual particle density expression is recovered as in reference \cite{lipatov_hybrid_2002}. As a result, inclusion of magnetic compression effectively leads to a 1.5D-3V PIC formulation.
\begin{equation}
n_\alpha(x,t)=N_R^\alpha\int{f_\alpha(x,\mathbf{v},t)}d^3v
\label{zerothMom}
\end{equation}
\begin{equation}
n_\alpha(x,t)=\frac{N_R^\alpha}{N_{SP}^\alpha}\frac{1}{A_0}\sum_{i=1}^{N_C^\alpha}{a_i^\alpha \frac{B_x(x_i^\alpha)}{B_{x0}}\delta(x-x_i^\alpha)} \hspace{1cm} \mathrm{[m^{-3}]}
\label{ne_compressed}
\end{equation}
\subsection{Finite boundaries }
\label{BC}
When particles exceed the boundaries of the computational domain, they are re-introduced back into the domain according to a set of rules. These rules determine the state of the computational particle on the next time step in the simulation. The state of each computational particle is completely defined by its (1) position $x_i$, (2) velocity vector $\mathbf{v}_i$ and (3) weight $a_i$.  Depending on how the particle state is updated after it has left the computational domain, periodic, reflecting or fully-absorbing boundary conditions for the particles can be implemented. For example, to model a system with no physical boundaries, a simple periodic boundary condition for the computational particles can be used where only the position state is updated. To model a system with finite boundaries, fully-absorbing boundary conditions for the particles are required where all the particle states are updated.

\subsection{Volumetric particle sources}
In the present study, the rules which update the state of the computational particles upon re-injection are chosen to model volumetric plasma sources such as: Neutral Beam Injection (NBI), isotropic warm-plasma sources or a combination of them.  

To update the computational particle position $x_i$ and velocity vector $\mathbf{v}_i$, the PDF of the particle source needs to be specified. For example to model NBI, the PDF of the beam needs to be specified using the following parameters: (1) beam injection location and (2) beam width in physical space, (2) beam energy, (3) beam temperature and (4) beam pitch angle. Setting the beam energy to zero and setting the temperature to a suitable value allows modelling a warm plasma source. Using these parameters, positions $x_i$ and velocity vectors $\mathbf{v}_i$ are randomly sampled from the plasma source PDF using a Box-Muller or Metropolis-Hastings (MH) algorithm \cite{Metropolis1953}.

In the strategy employed to update the computational particle weight $a_i$, it is necessary to distinguish between "real", "super" and "computational" particles. The PIC code employs a fixed number of computational particles $N_C$; that is, the number of memory locations updated in each PIC cycle is fixed. However, to allow for a time-dependent real particle density driven by volumetric particle sources and sinks, each computational particle must have a time-dependent weight. Since the total number of computational particles $N_C $ at any time step is fixed; the flux of computational particles out of the domain is equal to the flux into the domain. Hence, the new particle weight $a_i$ is calculated by dividing the requested particle injection rate by the influx of computational particles. In general, the requested particle injection rate can be time-dependent. In this strategy, the computational particle weight $a_i$ can change \textit{only} if it crosses the boundaries of the computational domain and is re-injected into the simulation.

\section{Particle and momentum transport in the paraxial limit}
\label{GBal}
As stated in the introduction (section \ref{Introduction}), mirror physics forms the basis of plasma systems where, in addition to velocity space, the physical dimension along the magnetic field must be resolved to extract the desired physics. An important step in the development of the 1D-3V hybrid PIC for open magnetic geometries is to confirm that the effect of spatially-dependent magnetic fields is correctly described. This can be achieved by testing conservation of momentum along the computational domain.

The purpose of this section is to present the particle and momentum transport equations for a plasma in the so-called paraxial limit \cite{newcomb_1981,newcomb_1985,Cohen1986}. In this framework, moments of the Vlasov equation are calculated in the limits of small inverse aspect ratio $ R/L \ll 1 $ and small gyro-radius $r_L/R \ll 1$ where $L$ and $R$ are the axial and radial scale lengths respectively and $r_L$ is the gyro-radius. Important assumptions required to produce the transport equations in the paraxial limit are also presented.
\subsection{Assumptions}
\subsubsection{The magnetic field in the paraxial limit}
$\\$
The externally applied magnetic field $\mathbf{B}_0$ is described in cylindrical coordinates $(r, \phi, x)$ by Eq. \ref{B_vector}, where $x$ is the coordinate along the length of the plasma system. The azimuthal magnetic field components $B_\phi$ is neglected. The radial magnetic component $B_r$  (Eq. \ref{B_r}) is obtained by setting $\nabla \cdot \mathbf{B}_0 = 0$  and assuming that $B_x$  varies only along the $x$  coordinate. This approximation is well suited for long-thin magnetized plasma columns and has the property that $B_x \gg B_r$. This implies that the angle $\gamma$ between the magnetic field vector $\mathbf{B}_0$ and $\hat{x}$ is small ($\gamma\ll1$) such that $\mathrm{sin} \gamma \approx \gamma$ , $\mathrm{cos} \gamma \approx 1$ , $\mathrm{tan}\gamma\approx\gamma$ and $B_x\approx |\mathbf{B}_0|=B$.
 \begin{equation}
\mathbf{B}_0 = B_x\mathbf{\hat{x}} + B_r\mathbf{\hat{r}}
\label{B_vector}\\
\end{equation}
\begin{equation}
 B_r = -\frac{r}{2}\frac{\partial B_x}{\partial{x}}
\label{B_r}
\end{equation}
\subsubsection{Magnetic field aligned coordinate system}
$\\$ 
A coordinate system aligned with the magnetic field can be produced whose unit vectors are given by Eq. \ref{s_hat} and Eq. \ref{n_hat}, where $\gamma$  is the angle between the magnetic field vector and $\mathbf{\hat{x}}$. The unit vector $\hat{s}$ is tangent to the magnetic field while $\hat{n}$ is normal to both $\hat{s}$ and $\hat{\phi}$. In this way, the unit vectors $\hat{s}, \hat{n}, \hat{\phi}$ form an orthogonal set and thus a coordinate system aligned with the magnetic field vector at every point along the plasma.
\begin{eqnarray}
\mathbf{\hat{s}}=+\cos\gamma\mathbf{\hat{x}} + \sin\gamma\mathbf{\hat{r}}
\label{s_hat}\\
\mathbf{\hat{n}}=-\sin\gamma\mathbf{\hat{x}}  +\cos\gamma\mathbf{\hat{r}}
\label{n_hat}
\end{eqnarray}Moreover, the differential operators in cylindrical coordinates expressed in terms of the field-aligned coordinates are given by Eq. \ref{d_dx} and Eq. \ref{d_dgamma}.
\begin{eqnarray}
\frac{\partial}{\partial x}=\cos\gamma\frac{\partial}{\partial s}-\frac{\sin\gamma}{s}\frac{\partial}{\partial\gamma}
\label{d_dx}\\
\frac{\partial}{\partial r}=\sin\gamma\frac{\partial}{\partial s}+\frac{\cos\gamma}{s}\frac{\partial}{\partial\gamma}
\label{d_dgamma}
\end{eqnarray}
\subsubsection{Radial transport constraint}
$\\$
In cylindrical coordinates, the bulk plasma flow is described by Eq. \ref{u_CylCoords}, where $\hat{x}$ is the unit vector along the length of the plasma and $\mathbf{\hat{r}}$ the radial unit vector. Assuming that the parallel transport far exceeds the radial transport under conditions of strong magnetization, the radial transport can be constrained by declaring $\mathbf{u}\cdot\hat{n}=0$  which leads to Eq. \ref{RadTransConstraint}  and hereafter referred to as the "radial transport constraint" \cite{wright_experimentally_2017}.
\begin{equation}
\mathbf{u}=u_x\mathbf{\hat{x}}+u_r\mathbf{\hat{r}}+u_\phi\mathbf{\hat{\phi}}
\label{u_CylCoords}
\end{equation}\begin{equation}
u_r = u_x \frac{B_r}{B_x}
\label{RadTransConstraint}
\end{equation}
\subsection{Particle and momentum transport}
The zeroth and first moments of the Vlasov equation lead to the \textit{particle} and \textit{momentum} transport equations for the species $\alpha$  presented in Eq. \ref{GparticleTrans} and Eq. \ref{GmomTrans} and taken from Ref.(\cite{Meier2012}).

\begin{eqnarray}
\frac{\partial n_\alpha}{\partial t} + \nabla \cdot (n\mathbf{u_\alpha}) = G_\alpha
\label{GparticleTrans}\\
\frac{\partial}{\partial t} (m_\alpha n_\alpha \mathbf{u_\alpha}) + \nabla \cdot (m_\alpha n_\alpha \mathbf{u_\alpha u_\alpha} + \mathbb{P_\alpha}) =q_\alpha n_\alpha (\mathbf{E} + \mathbf{u_\alpha} \times \mathbf{B})
\label{GmomTrans}
\end{eqnarray}Assuming that inertia is dominated by ions and using quasi-neutrality ($n=n_e=n_i$), the single-fluid \textit{momentum} transport equation is given by Eq. \ref{singleFluidMomTrans}, where $M= m_i+m_e$, $\mathbf{u}$ is the plasma bulk velocity given by Eq. \ref{SingleFluidFlow} and the plasma pressure $\mathbb{P}$ is given by Eq. \ref{SingleFluidPressure}. 
\begin{equation}
\frac{\partial}{\partial t}(Mn\mathbf{u}) + \nabla \cdot (M n \mathbf{u u}) =-\nabla \cdot \mathbb{P} + \mathbf{J} \times \mathbf{B}
\label{singleFluidMomTrans}
\end{equation}\begin{equation}
\mathbf{u}=\frac{m_i\mathbf{u}_i + m_e\mathbf{u}_e}{m_i + m_e}
\label{SingleFluidFlow}
\end{equation}\begin{equation}
\mathbb{P}=\mathbb{P}_i+\mathbb{P}_e
\label{SingleFluidPressure}
\end{equation}To proceed,  divergence operators in \ref{singleFluidMomTrans} are expanded in cylindrical coordinates and then transformed to the field-aligned coordinate system using Eq. \ref{d_dx} and Eq. \ref{d_dgamma} and the unit vectors given in Eq. \ref{n_hat} and Eq. \ref{s_hat}. Moreover, the pressure tensor $\mathbb{P_\alpha}$ is given in terms of the field-aligned coordinate system as shown in Eq. \ref{PressureTensor}. The radial transport constraint (Eq. \ref{RadTransConstraint}) is applied and the magnetic field in the paraxial limit is used. This leads to the \textit{particle} and single-fluid parallel \textit{momentum} transport equations in the paraxial limit presented in Eq. \ref{ParaxialParticleTrans} and Eq. \ref{ParaxialSingleFluidMomTrans} respectively, where $s$ is the coordinate along the magnetic flux. \begin{equation}
\mathbb{P_\alpha} = P_{\alpha\parallel} \mathbf{\hat{s}\hat{s}} + P_{\alpha \perp} \mathbf{\hat{n} \hat{n}} +  P_{\alpha \perp} \mathbf{\hat{\phi}	\hat{\phi}}
\label{PressureTensor}
\end{equation}\begin{equation}
\frac{\partial n_\alpha}{\partial t} +B \frac{\partial}{\partial s}\bigg( \frac{n_\alpha u_{\parallel\alpha}}{B} \bigg) = G_\alpha
\label{ParaxialParticleTrans}
\end{equation}\begin{equation}
\frac{\partial}{\partial t} (Mnu_\parallel) + B \frac{\partial}{\partial s} \bigg( \frac{Mnu_\parallel u_\parallel}{B} \bigg) = - \frac{\partial P_\parallel}{\partial s} - \frac{P_\perp - P_\parallel}{B} \frac{\partial B}{\partial s}
\label{ParaxialSingleFluidMomTrans}
\end{equation}
Eq. \ref{ParaxialSingleFluidMomTrans} is a general statement of momentum conservation in one dimension in physical space (along the magnetic flux) and includes the effect of a \textit{non-uniform} magnetic field. As the plasma flows into regions of higher magnetic field, it experiences magnetic compression due to the reduction in cross-sectional area. It is worthwhile to note that in steady-state and in the absence of plasma flow, Eq. \ref{ParaxialSingleFluidMomTrans} reduces to the classic mirror force balance equation as shown in Eq. \ref{MirrorBalance} and taken from reference \cite{goldston_introduction_2020}.

\begin{equation}
 \frac{\partial P_\parallel}{\partial s} + \frac{P_\perp - P_\parallel}{B} \frac{\partial B}{\partial s}=0
 \label{MirrorBalance}
\end{equation}
\subsubsection{Distribution function}
$\\$
It is important to note that in all the transport equations presented above, no assumption is made about the form of the distribution function. All terms represent integrals of an \textit{unspecified} distribution function. In general, each transport equation is coupled to the next higher order transport equation and leads to the problem of fluid closure. To provide closure, assumptions are made on the form of the distribution function as in the Braginskii fluid plasma formalism \cite{braginskii_transport_1965,fitzpatrick_plasma_2014} where the distribution function is explicitly assumed to be a Maxwellian to first order with first-order non-Maxwellian corrections. In the present work, the form of the distribution function is not imposed as this is calculated self-consistently by the hybrid PIC code and the transport equations herein presented, in particular (Eq. \ref{ParaxialSingleFluidMomTrans}), are specifically used to validate the results produced by the PIC code. More specifically, using the particle positions and velocities evolved by the PIC code, the moments of the distribution function ($n, u_\parallel, P_\parallel, P_\perp$) are calculated as a function of physical space and each term in Eq. \ref{ParaxialSingleFluidMomTrans} is evaluated to test conservation of momentum and thereby validate the new physics added to the code:  (1) non-uniform magnetic fields, (2) finite boundaries and (3) volumetric sources.

\section{Initial conditions for the simulation}
\label{InitialConditions}
In the present 1D-3V calculation, the state of each computational particle is defined by its: (1) position $x_i$, (2) velocity vector $\mathbf{v}_i$ and (3) weight $a_i$, Prior to evolving the state of the computational particles with the PIC code, the initial state of the entire set of particles needs to be specified. In the present study, the initial distribution of particles states is chosen so as to satisfy the paraxial mirror force equation (Eq. \ref{MirrorBalance}). In other words, the initial state of the plasma described by the PIC is close to a state of equilibrium or force balance. 

The general procedure for selecting a suitable initial condition for the entire particle set is the following:
		  
\begin{enumerate}
\item[Step-1:] Define the externally applied magnetic field profile $\mathbf{B}_0$ along the length of the computational domain. An example is shown in figure \ref{fig1}.
\item[Step-2:] Given the magnetic field profile $\mathbf{B}_0$ from Step-1 and suitable boundary conditions for the parallel and perpendicular pressure, solve the mirror equilibrium in the paraxial limit (Eq. \ref{MirrorBalance}) to obtain the parallel and perpendicular pressures profiles $P_\parallel$ and $P_\perp$ along the length of the plasma. This step is described in more detail in \ref{AppA}. A solution to Eq. \ref{MirrorBalance} is shown in figure \ref{fig2} using the magnetic field from figure \ref{fig1}. In this solution, the parallel pressure at the central cell $ P_{\parallel1}$ and mirror throats $ P_{\parallel2}$ are specified and the perpendicular pressures are given as $P_{\perp1} = P_{\parallel1}$ and $P_{\perp2} = P_{\parallel2}/3$.
\item[Step-3:] Construct a 4-D probability density function (PDF) whose moments are equivalent to the new pressure profiles $P_\parallel(x)$ and $P_\perp(x)$ obtained in step-2. The process to construct a suitable PDF is described in \ref{4D-pdf}.
\item[Step-4:] Apply a Metropolis-Hasting (MH) algorithm\cite{Metropolis1953, Hasting1970,Calderhead2014} to randomly extract positions $x_i$ and velocity vectors $\mathbf{v}_i$ from the 4-D PDF. This process creates a distribution of discrete particles consistent with the mirror equilibrium in the paraxial limit (Eq. \ref{MirrorBalance}).
\item[Step-5:] Set the particle positions $x_i$ and velocity vectors $\mathbf{v}_i$  from Step-4 as the initial condition for Hybrid PIC simulation.
\end{enumerate}

\subsection{Simulation condition and input data}
\label{simparsec}
The hybrid PIC code PRO++OMG is targeted to any system with open magnetic geometries where the physical dimension along the magnetic field needs to be retained such as in the SOL of fusion devices,  linear divertor simulators, mirror devices and plasma thrusters. In the present study, PRO++OMG is applied to a magnetic mirror geometry. The magnetic field profile and nominal plasma conditions are taken from the Wisconsin HTS Axisymmetric Mirror (WHAM) device\cite{Anderson_FST_2021}. The magnetic field profile is presented in figure \ref{fig1}. The various parameters used in the simulation are tabulated in the Table-\ref{arttype1} below:

\begin{table}[H]\label{SimPar}
\caption{\label{arttype1} Simulation parameters for Hybrid PIC simulation}
\footnotesize\rm
\begin{tabular*}{\textwidth}{@{}l*{15}{@{\extracolsep{0pt plus12pt}}l}}
\br
Simulation parameters  & \hspace{3cm} Physical value\\
\mr
\verb"Length of the domain" $L$&\hspace{3cm} $3.19\hspace{1mm} [\mathrm{m}]$\\
\verb"Coil to coil distance" $L_m$&\hspace{3cm} $2.0\hspace{1mm}[\mathrm{m}]$\\
\verb"Peak magnetic field" $B_2$&\hspace{3cm} $17\hspace{1mm}[\mathrm{T}] $\\
\verb"Central cell magnetic field" $B_1$&\hspace{3cm} $0.6\hspace{1mm}[\mathrm{T}]$\\
\verb"Central cell plasma radius" $R_{1}$&\hspace{3cm} $0.1\hspace{1mm}[\mathrm{m}]$ \\
\verb"Electron temperature" $T_e$&\hspace{3cm} $1\hspace{1mm}[\mathrm{keV}]$\\
\verb"Ion mass number" $A_\alpha$&\hspace{3cm} $1\hspace{1mm}[\mathrm{AMU}]$\\
\verb"Ion charge number" $Z_\alpha$&\hspace{3cm} $+1$\\
\verb"Ion density fraction" &\hspace{3cm} $1$\\
\verb"Number of cells"&\hspace{3cm} $500$\\
\verb"Number of computational particles per cells"&\hspace{3cm} $5000$\\
\br
\end{tabular*}
\end{table}

\begin{figure}[H]
\center
			\textbf{(a) \hspace{7cm}(b)}\par\medskip
            \includegraphics[width=0.53\textwidth]{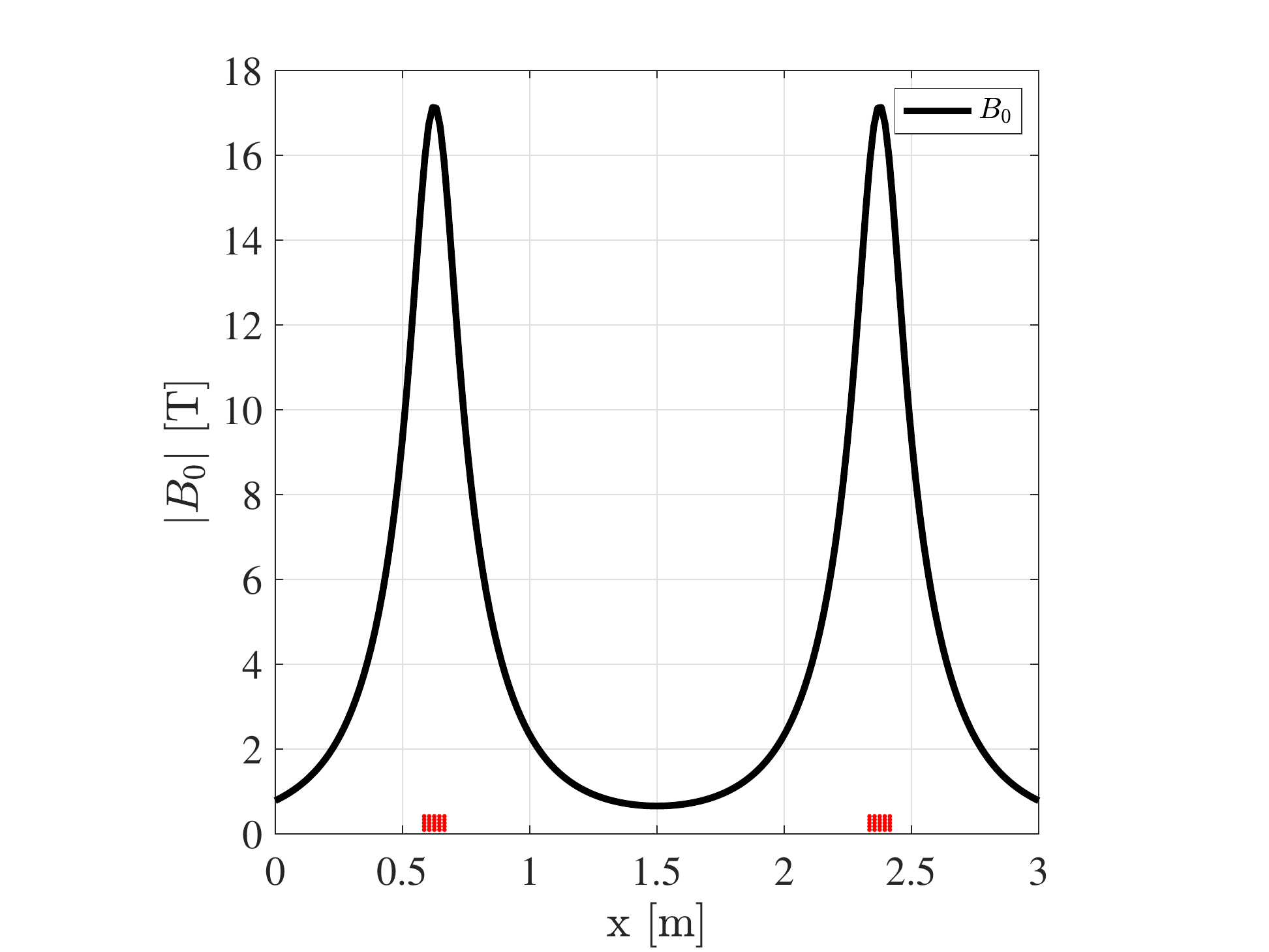} 
            \hspace{-1.5cm}\includegraphics[width=0.53\textwidth]{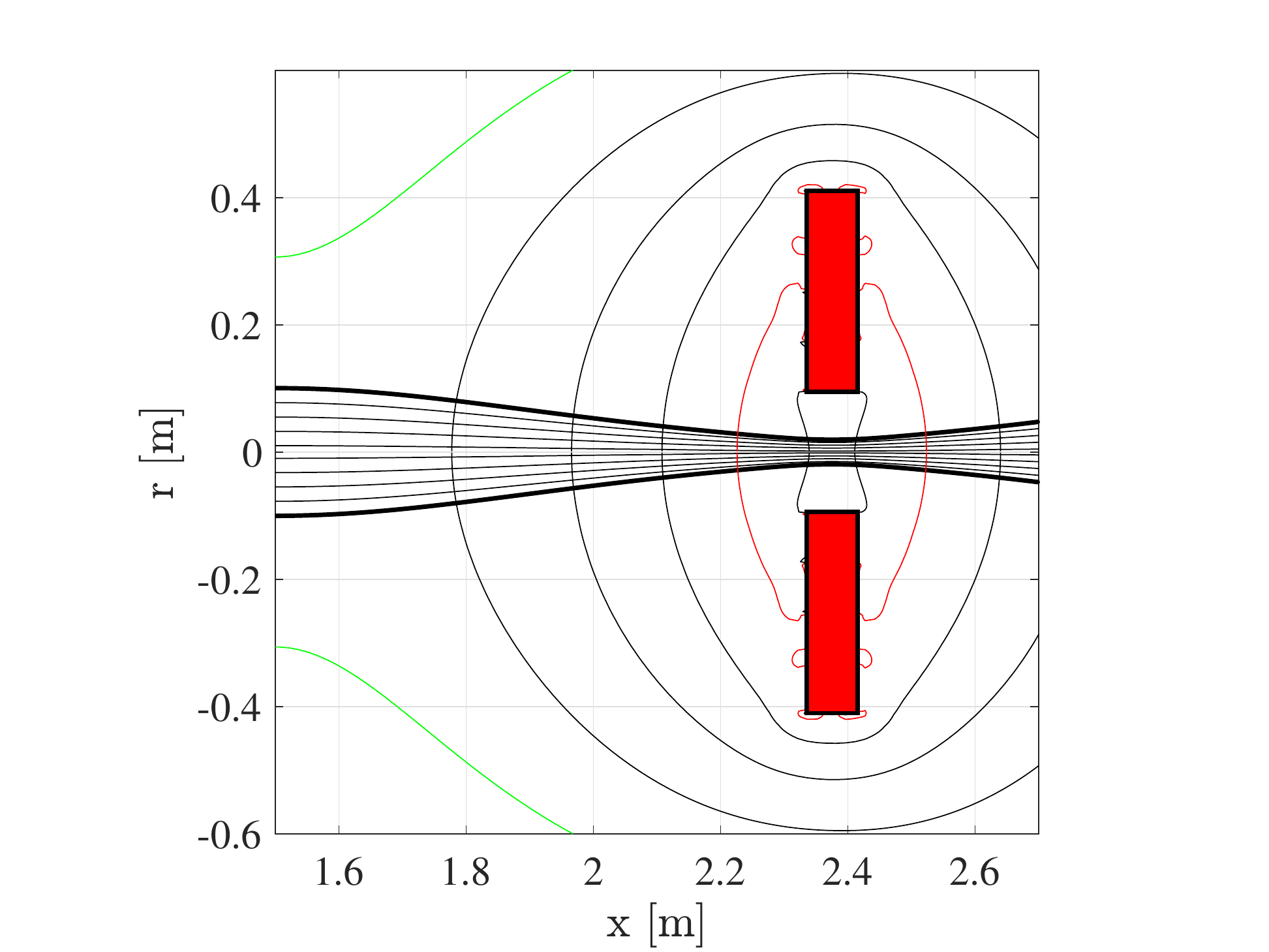} 
            \caption{(a) Axial magnetic field profile based on a pair of $17\hspace{1mm}\mathrm{T}$ magnets separated by a distance of 2 meters. The location of the coils is represented by the red squares. The magnetic field strength at the central cell  is $0.6\hspace{1mm}\mathrm{T}$.  (b) Magnetic flux lines. Red rectangular box shows the position of the coils.}  
            \label{fig1}
\end{figure} 

\section{Results}
\subsection{Plasma profiles:}
\label{PP}
To start the 1D-3V hybrid PIC calculation, the initial state of the particles needs to be selected as described in section \ref{InitialConditions}. This requires the selection of boundary conditions for the plasma pressure profiles at the central cell and mirror throats as described in table \ref{arttype2}. Near the edges of the computational domain in the so-called exhaust regions, the fully-absorbing particle boundary conditions render the paraxial mirror equilibrium invalid, hence the pressure profiles are truncated at the mirror throats. Beyond the mirror throats, the pressure profiles are described with hyperbolic tangent functions to model the decay of the plasma and temperature in the exhaust regions. 

\begin{table}[H]\label{BCforStartUp}
\caption{\label{arttype2} Input parameters for calculating paraxial mirror equilibrium}
\footnotesize\rm
\begin{tabular*}{\textwidth}{@{}l*{15}{@{\extracolsep{0pt plus12pt}}l}}
\br
Simulation parameters  & \hspace{3cm} Physical value\\
\mr
\verb"Central cell plasma density" $n_{e1}$&\hspace{3cm}  $5\times 10^{19} \hspace{1mm}[\mathrm{m}^{-3}]$\\
\verb"Central cell electrons temperature" $T_e$&\hspace{3cm} $1\hspace{1mm}[\mathrm{keV}]$\\
\verb"Central cell parallel ion pressure" $P_{\parallel1}$&\hspace{3cm} $n_{e1}T_{e1}\hspace{1mm} [\mathrm{Pa}]$\\
\verb"Central cell perpendicular ion pressure" $P_{\perp1}$&\hspace{3cm} $P_{\parallel1}\hspace{1mm} [\mathrm{Pa}]$\\
\verb"Throat parallel ion pressure" $P_{\parallel2}$&\hspace{3cm} $P_{\parallel1}/6\hspace{1mm} [\mathrm{Pa}]$\\
\verb"Throat perpendicular ion pressure" $P_{\perp2}$&\hspace{3cm} $P_{\parallel1}/3\hspace{1mm} [\mathrm{Pa}]$\\
\br
\end{tabular*}
\end{table}

\begin{figure}[H]
\center
            \includegraphics[width=0.85\textwidth]{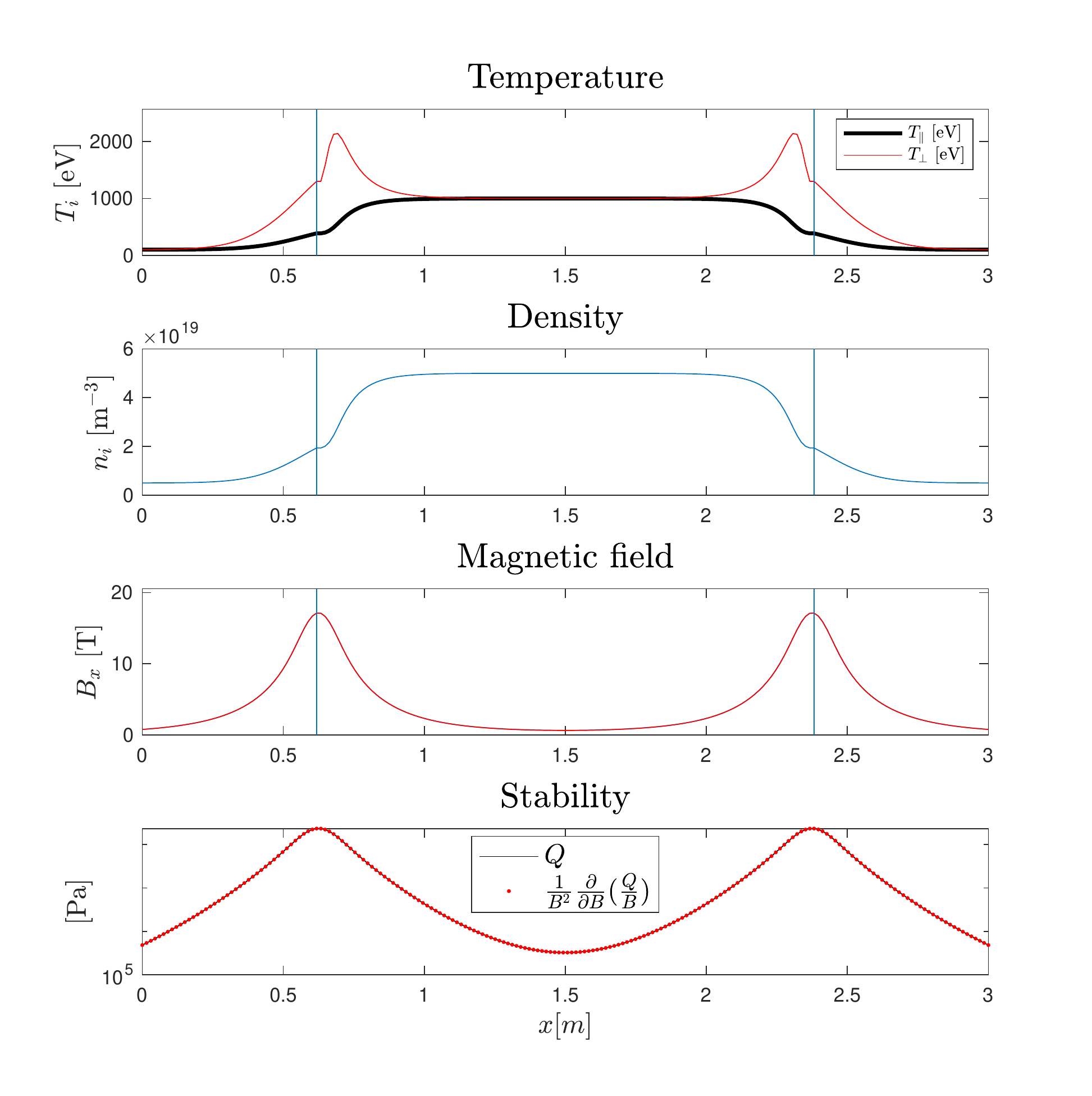} 
            \caption{Plasma profiles obtained using the parallel and perpendicular force balance equations as described in \ref{EqProfiles}. These profiles are used in PRO++-OMG as initial condition to be loaded at $t=0$ as an external files. }  
            \label{fig2}
\end{figure}

Using the parameters from table \ref{arttype2} and the procedure outlined in \ref{AppA}, the equilibrium plasma profiles are shown in figure \ref{fig2}.  Near the mirror throats, the ion temperature acquires an anisotropic nature. This is expected as pressure anisotropy is required to balance the parallel pressure gradient force as demonstrated by Eq. \ref{MirrorBalance}. The bottom panel evaluates the term $Q$ discussed in \ref{AppA} which is related to stable equilibrium of the profiles. The solution shows that $Q>0$  in all regions of the domain. Equilibrium solutions that do not satisfy $Q>0$ are rejected as recommended in reference\cite{Cohen1986}.

In the present case, the plasma system is evolved up to 3000 gyro-periods ($t=3000 \Omega_c^{-1}$) and fueled with a warm isotropic (Maxwellian) plasma source injected at the central cell. The plasma source has a Gaussian distribution in physical space with a spread of $L/10$  about the center of the cell. The gyro-period is calculated based on the magnetic field in the central cell which is set to 0.6 T. This corresponds to a total simulation time of 0.33 ms. The isotropic warm plasma source is set to have a temperature of 1 $\mathrm{keV}$ and injects ions at a rate which maintains the central density equal to $5\times 10^{19}\hspace{1mm} [\mathrm{m^{-3}]}$. The addition of a warm isotropic plasma source implies that the collision-less plasma system will evolve to be a combination of a mirror-trapped/loss-cone distribution and an isotropic Maxwell-Boltzmann distribution. This is demonstrated in section \ref{LossConeDistribution}. 

\begin{figure}[H]
\center
            \includegraphics[width=\textwidth]{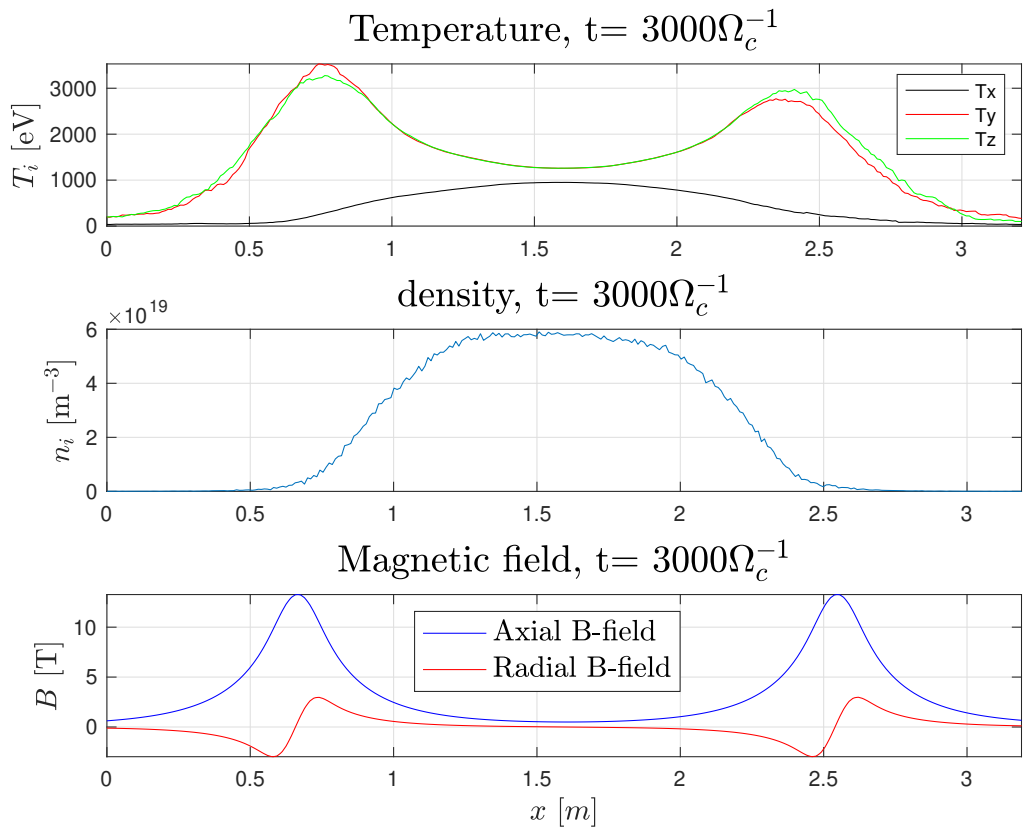} 
            \caption{Plasma profiles for ion temperatures $T_x,T_y$ and $T_z$ along $\hat{x}, \hat{y}$ and $\hat{z}$ directions respectively, ion density $n$, axial and radial magnetic field   obtained from the collision-less simulations of mirror plasma at $t=3000 \Omega_c^{-1}$. These profiles obtained at $t=3000$ ion-gyro periods are very much similar to the equilibrium profiles shown in Fig.(\ref{fig2}) loaded as an initial condition at $t=0$.}  
            \label{fig3}
\end{figure}

Using the computational particle positions, velocity vectors and weights evolved by the PIC code, the moments of the distribution function are calculated to obtain the plasma density, ion flux and ion pressure tensor profiles along the length of the plasma and as a function of time. The axial variation of the ion temperature (parallel and perpendicular) and plasma density are presented in Figure \ref{fig3} at $t=3000 \Omega_c^{-1}$. The results show that both the temperature and density profiles evolve to a state that is qualitatively similar to the equilibrium profiles used to initialize the distribution at the start of the simulation (Figure \ref{fig2}); namely, the temperature develops an anisotropic nature near the mirror throats. This suggests that the plasma system has evolved while maintaining force balance. This is confirmed in section \ref{ConservationOfMomentum}.

It is worth noting that for the present conditions, Coulomb collisions are expected to play a role only after several thousands of ion-gyro periods. Using equation 1 from reference\cite{caneses_effect_2020}, the mean time for 90 degree scattering a thermal proton on a background proton plasma with a mean temperature 1.5 keV and density $5\times 10^{19}\hspace{1mm}[\mathrm{m^{-3}]}$ is about 0.33 ms. Hence at 3000 gyro-periods which in the present case correspond to about 0.3 ms, the dynamics of the plasma system can still be considered quasi-collision-less; nevertheless, 3000 gyro-periods is sufficient time to allow evolution of the ion trajectories in phase space and hence the evolution of the ion PDF. Therefore, in order to proceed to several tens of thousands gyro-periods and into the tens of milliseconds range, the addition of Coulomb collisions becomes imperative.  This will be the subject of another paper. In the discussion, a brief description of the planned Coulomb collision operator is presented.

\subsection{Conservation of Momentum }
\label{ConservationOfMomentum}
\begin{figure}[H]
\center
            \includegraphics[width=0.5\linewidth]{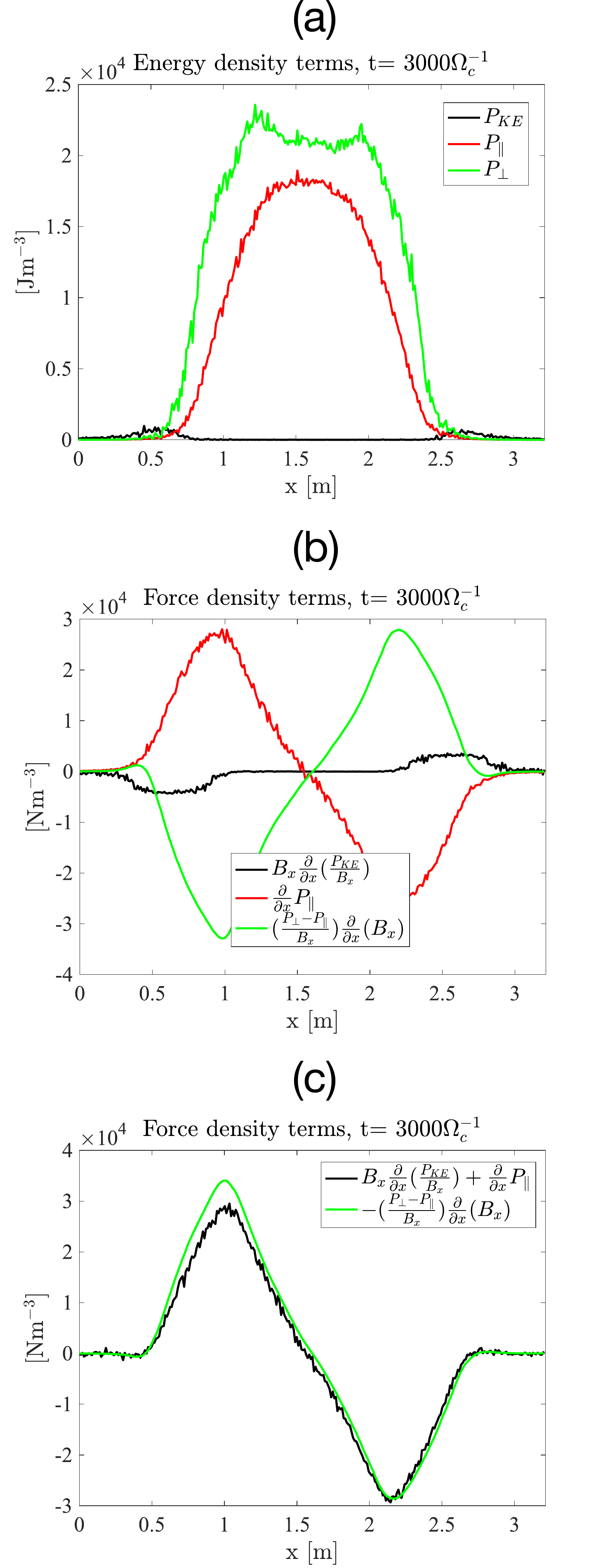} 
            \caption{Comparison of simulation data with the momentum transport equation in the paraxial limit. (a) Axial variation of individual energy density terms- Parallel drift kinetic energy $Mnu_\parallel^2$ (solid black curve ), Perpendicular $P_\perp$and parallel $P_\parallel$ pressures (green and red solid curves) ; (b) corresponding derivative of energy density terms; and (c) Comparison of analytic value (solid black curve) of mirror force with that obtained from the simulation data (solid green curve). }  
            \label{fig4}
\end{figure}

The process of calculating the moments of the ion PDF using the particle positions in phase space naturally leads to the evaluation of conservation of momentum or force balance. Specifically, using the plasma density, plasma flux and ion pressure tensor evolved by the PIC code, each term of the momentum transport equation in the paraxial limit (Eq. \ref{ParaxialSingleFluidMomTrans}) can be evaluated as a function of time and space to test conservation of momentum. This process has been carried out in the present study using the simulation results at $t=3000 \Omega_c^{-1}$ and presented in figure \ref{fig4}. 

The spatial variation of the ion pressure tensor is shown in figure \ref{fig4}a, where the term $P_{KE}$ represents the term associated with the kinetic energy density given by $P_{KE} = Mnu_{\parallel}u_{\parallel}$. The dominant term is the perpendicular pressure leading to an anisotropic ion pressure; however, near the throats, as the plasma exhaust begins to accelerate (see Figure \ref{fig5}b), the kinetic energy density of the plasma $P_{KE}$  increases. Based on these spatially-varying pressure profiles, each term in paraxial mirror momentum transport equation (Eq. \ref{ParaxialSingleFluidMomTrans}) is evaluated and presented in figure \ref{fig4}b. The time-dependent term is not included as it is vanishingly small during the approach to steady state. The red line represents the parallel pressure gradient, the green line the magnetic mirror force and the black line the force associated with accelerating the plasma through the mirror. The data indicates that the parallel gradient force (red line) is balanced mostly by the magnetic mirror force (green line). To confirm conservation of momentum, Figure \ref{fig4}c compares two terms of Eq. \ref{ParaxialSingleFluidMomTrans} and indicates that indeed momentum is conserved. In other words, the plasma system at 3000 gyro-periods is in a state of force balance or equilibrium and consistent with the momentum transport equation in the paraxial limit.  

\subsection{Self-consistent ambipolar electric field}
\begin{figure}[H]
\center
            \includegraphics[width=0.75\linewidth]{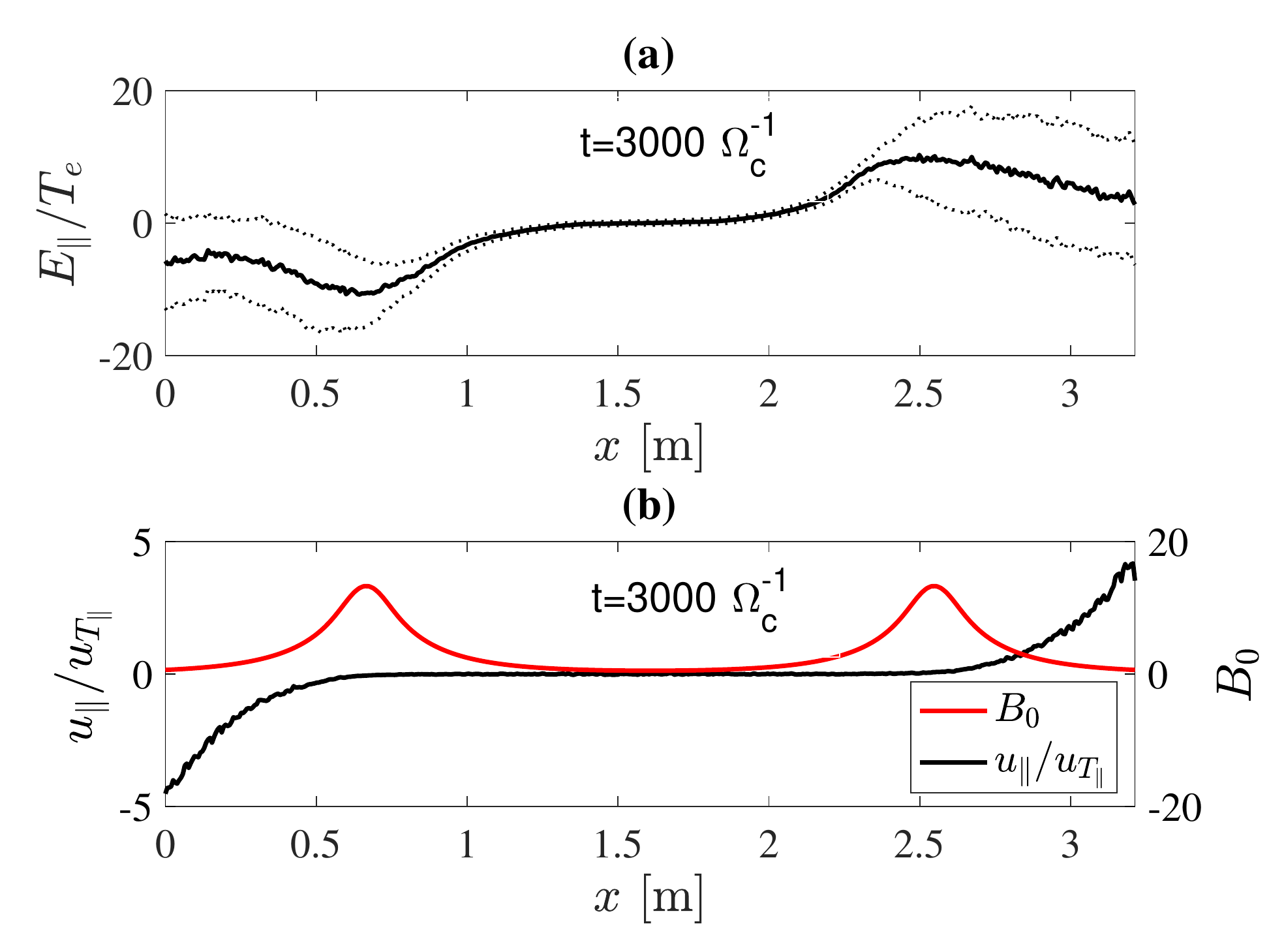} 
            \caption{(a) Self-consistent parallel electric field $E_\parallel$ profile at $t=3000 \Omega_c^{-1}$along the magnetic flux lines normalized to the electron temperature $T_e$. The dashed black curve represent the standard deviation the electric field (b) Parallel ion drift velocity profile normalized with the average ion sound speed at $t=3000 \Omega_c^{-1}$ (solid black curve) superposed over the magnetic field profile (red curve).}  
            \label{fig7}
\end{figure}

As the PIC code evolves the particle positions in phase space, the parallel electric field $E_\parallel$ is self-consistently calculated using the generalized Ohm's law presented in Eq. \ref{GOHM}. This electric field in turn, affects the parallel plasma transport. Figure \ref{fig7} presents the parallel electric field along the length of the plasma at $t=3000 \Omega_c^{-1}$. Moreover, the parallel plasma drift velocity is shown for completeness indicating the presence of plasma flows of several ion thermal velocities towards the boundaries.

The electric field $E_\parallel$ is approximately flat in the central cell region and increases to a maximum near the mirror throats. The sign of the electric field indicates that it expels ions and confines electrons in the central cell. The presence of this parallel electric field has important consequences on the formation of mirror-trapped distributions in the form of so-called Yushmanov potentials\cite{Ryutov1988}. This is discussed further in the next section. In the exhaust region, the noise level of the electric field increases as illustrated by the error bars. These fluctuations are most likely caused by the poor statistics in this region as the plasma density drops sharply in the exhaust region (see Figure \ref{fig3}). To improve the statistics in this region, there are two actions that can be taken. The first one is to increase the particle interpolation scheme. Currently, PRO++ OMG utilizes a second-order interpolation scheme and implementing a third order scheme could help reduce noise\cite{Shalaby_2017}. In addition, a particle splitting scheme can be applied in the exhaust region to improve statistics and thus the noise levels on the electric field calculation (\cite{SMETS2021107666, PFEIFFER20159})

It is worth noting that resolving the parallel electric field in a self-consistent manner is required to correctly describe the parallel transport of ions along the magnetic flux. This becomes increasingly important in devices like WHAM \cite{Anderson_FST_2021} and MPEX\cite{caneses_effect_2020} where radio-frequency plasma heating can produce mirror-trapped fast particles. As the density of trapped fast particles increases, the parallel electric field changes and affects the parallel transport of the bulk plasma. An important example of this is the use NBI to create a sloshing ion population in a mirror device which improves the confinement of the background warm plasma. On the other hand, in devices such as MPEX, care must be taken to prevent the formation of mirror-trapped particles during RF heating which reduce the parallel transport of particles towards the target region\cite{caneses_effect_2020}. 

\subsection{Loss cone ion distribution function}
\label{LossConeDistribution}
The conditions for plasma confinement in a mirror configuration in the presence of an ambipolar potential is determined by the condition presented in Eq. \ref{MirrorConfinementCondition} and obtained from references\cite{Kaufman1956,BenDaniel_1961, Yushmanov1966, Ryutov1988}. In this expression, $v_\perp$ and $v_\parallel$ are the perpendicular and parallel ion velocities at the center of the mirror region respectively, $R_m = B_m/B_1$ is the mirror ratio; $e$ is the electronic charge and $\Delta \phi$ is the electric potential between the central cell and the mirror throat. The confinement region/loss cone boundary in the presence of an ambipolar potential becomes a hyperboloid in velocity space coordinates.
\begin{equation}
\label{MirrorConfinementCondition}
v_{\perp}^2 (1-\frac{1}{R_m}) -\frac{v_\parallel^2}{R_m} > \frac{2e\Delta \phi}{MR_m}
\end{equation}
The ion velocity space distribution in the central cell is presented in figure \ref{fig7}(a, b \& c) at $t=3000 \Omega_c^{-1}$ . The dotted black lines represent the loss cone boundary in the \textit{absence} of an electric potential. These straight lines have a pitch angle relative to the $v_\parallel$  coordinate given by $\sin ^{-1}\sqrt{1/R_m}$ . The black solid lines which trace hyperbolas represent the confinement boundary given by Eq. \ref{MirrorConfinementCondition} where $\Delta\phi$ is obtained by integrating the parallel electric field presented in Figure \ref{fig7}a from the central cell to the mirror throat. Ions with energies less than $e\Delta \phi /R_m$  exist below the confinement boundary and are lost from the mirror region. On the other, ions with energies greater than $e\Delta \phi /R$  can exist above the confinement boundary  given by Eq. \ref{MirrorConfinementCondition} and thus have a greater probability to remain confined.

\begin{figure}[H]
\center
        \includegraphics[width=1.15\textwidth]{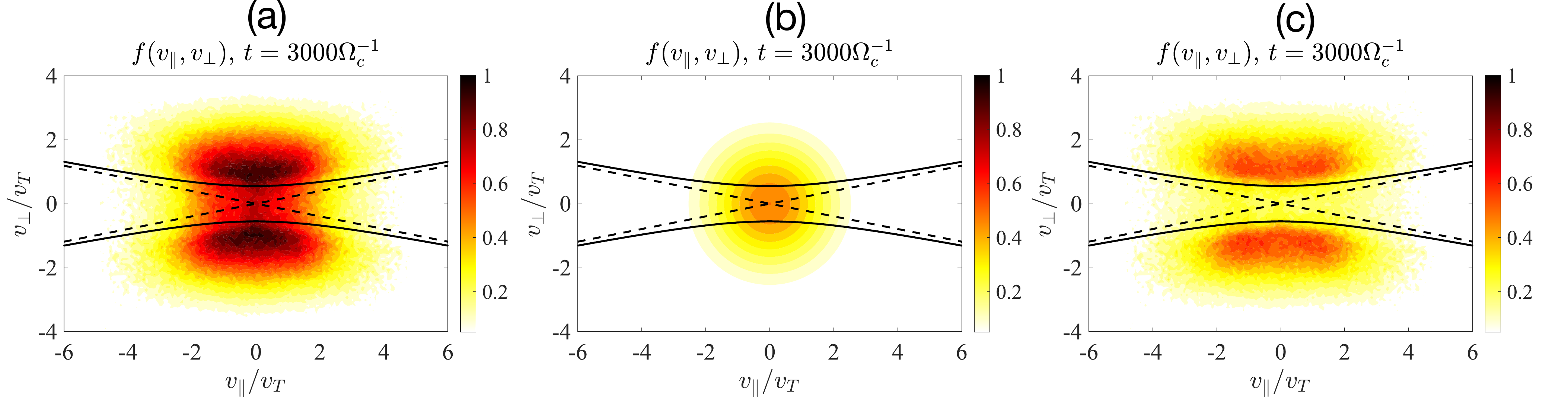} 
        \caption{Velocity distribution of ions in mirror plasma configuration: (a) Total distribution of ions in presence of an isotropic particle source (b) Particle source distribution, and (c) Loss cone distribution of ions in absence of any particle source. The solid and dashed black curve in figures are boundaries of the confinement regions with and without ambipolar potential respectively. }  
        \label{fig5}
\end{figure} 

Figure \ref{fig5}a presents the total ion velocity distribution obtained directly from the PIC calculation. As discussed in section \ref{PP}, the plasma system is fueled with a warm isotropic plasma source injected at the central cell with a temperature of 1 keV. The velocity space distribution of this plasma source is shown in Figure \ref{fig5}b. In fact, subtracting the plasma source velocity distribution (figure \ref{fig5}b) from the total distribution presented in figure \ref{fig5}a produces the distribution presented in figure \ref{fig5}c. This distribution is precisely a loss-cone distribution in the \textit{presence} of an electric potential. Regions of high particle density are accurately described by the the hyperbolas representing Eq. \ref{MirrorConfinementCondition}. This figure clearly demonstrates that the total ion velocity distribution in the central cell at $t=3000 \Omega_c^{-1}$ can be understood as a combination of a loss cone-distribution driven by mirror physics and an isotropic Maxwellian distribution driven by the warm plasma source.

\section{Discussion}
The development of Pro++ OMG has been motivated by the need to develop tools to model the evolution of the distribution function in plasma systems with open magnetic geometries in the presence of non-uniform magnetic fields, RF heating and collisional process. Examples of such systems include (1) the SOL in toroidal fusion confinement devices, (2) linear divertor simulator such as the MPEX under construction at ORNL for fusion PMI research, (3) mirror devices such as the Wisconsin HTS Axisymmetric Mirror (WHAM) under construction at the University of Wisconsin Madison, (4) plasma thrusters such as the VASIMR concept which uses Ion Cyclotron Resonance Heating (ICRH) to heat a low-temperature high-density plasma to increase thrust.

A key aspect of plasmas systems in open magnetic geometries is the need to retain/resolve the physics along the physical dimension associated with the magnetic field. Depending on the level of collisionality, the presence of non-uniform magnetic fields introduces velocity space anisotropies leading to mirror forces on the plasma and non-uniform plasma densities. Moreover, the use of RF plasma heating via resonant methods  and/or injection of energetic particles further drives velocity space anisotropies which in general becomes a function of space. The minimum theoretical model needed to capture all these effects is the Boltzmann transport equation with at least 1 dimension in physical space (1D) and at least two dimensions velocity space dimension (2D). In addition, source terms are required to account for plasma sources/sinks and collisional process. In the context of plasmas, the acceleration term in the Boltzmann transport equation becomes the Lorentz force and leads to the Vlasov equation. In the present work, a hybrid PIC approach is employed to solve the transport problem along 1 dimension in physical space (1D) and 3 dimensions in velocity space (3V).

\subsection{Initial state of the plasma system}
To evolve the hybrid PIC code, the initial state of each particle needs to be specified. The particle state in the current formulation is completely defined by: (1) position $x_i$, (2) velocity vector $\mathbf{v}_i$ and (3) weight $a_i$. This can be clearly seen in the discrete particle distribution function Eq. \ref{klimontovichCompressed}. To select the initial state of the entire set of particles, the general method described in section \ref{InitialConditions} is employed. A key step in this process is calculating the parallel $P_\parallel$ and perpendicular pressure $P_\perp$ profiles as a function of the spatial coordinate. This process amounts to the solution of the mirror equilibrium in the paraxial limit (Eq. \ref{MirrorBalance}). A simple method to compute these profiles is presented in \ref{AppA} where the required inputs are: (1) vacuum magnetic field profile and (2) four boundary conditions of the pressure profiles. The outputs are: (1) parallel pressure, (2) perpendicular pressure and (3) beta-corrected magnetic field. This method can be extended to problems with two dimensions (2D) in physical space by dividing the volume into nested flux surfaces. The paraxial mirror equilibrium is solved along each flux surface. Using the beta-corrected magnetic field profiles, the new nested flux surfaces are calculated and the 2D spatially-resolved pressure profiles can be assembled.

\subsection{Non-uniform magnetic fields}
In the present work, the focus has been on the correct implementation of non-uniform magnetic fields, finite fully-absorbing boundaries for the particles and volumetric plasma sources. The presence of non-uniform magnetic fields leads to "pinching" or "compression" of the plasma as particles move into regions of higher magnetic field. In the 1D-3V formulation, this effect is captured by a compression factor $B(x_i)/B_0$  in the discrete particle distribution function as presented in Eq. \ref{klimontovichCompressed} which effectively increases the density of super-particles as they enter regions of higher magnetic field. This term is essential in retaining the 2D nature of a cylindrically symmetric non-uniform magnetic field.

To demonstrate correct implementation of the non-uniform magnetic fields, the hybrid PIC is applied to a simple magnetic mirror geometry which features two finite fully-absorbing boundaries and a warm isotropic plasma source to sustain the plasma. The results of the code are evaluated using the momentum transport equation in the paraxial limit. This process consists of computing the moments of the distribution function based on the super-particle phase-space positions. The computed moments are used to evaluate each term of the momentum transport equation (Eq. \ref{ParaxialSingleFluidMomTrans}) to assess conservation of momentum along the length of the plasma system. The results presented in fig. \ref{fig4} describe the contribution of each force and demonstrates conservation of momentum along the length of the plasma system. The results show that the plasma exists in a state of force balance between the parallel pressure gradient and the mirror force. The plasma acceleration term becomes important only near the mirror throats where the plasma accelerates to super-sonic speeds (Figure \ref{fig5}b). 

In addition to evaluating conservation of momentum, the velocity distribution of the ions in the central cell is analyzed. It is well known in the literature\cite{Yushmanov1966,Ryutov1988} that the presence of an electric potential in a mirror plasma modifies the loss cone boundary from a cone to a hyperboloid. The exact shape of this boundary can be calculated with knowledge of the electric potential and the mirror ratio as shown in Eq. \ref{MirrorConfinementCondition}. The electric potential is self-consistently calculated in the hybrid PIC approach using the generalized Ohm's law (Eq. \ref{GOHM}) and is used to evaluate the loss boundary. The results presented in figure \ref{fig5} demonstrate that indeed the regions in velocity space with the highest ion density are those above the hyperbolic loss boundaries predicted by Eq. \ref{MirrorConfinementCondition}. In fact, the total ion distribution function can be understood as the combination of a loss-cone distribution and an isotropic Maxwellian distribution driven by the warm isotropic plasma source. This further demonstrate the correct implementation of non-uniform magnetic field physics in the present 1D-3V hybrid PIC.

\subsection{Coulomb collisions}
In the present work, the ion distribution function is evolved up to 3000 gyro-periods. Based on the central cell magnetic field, this corresponds to approximately 0.3 ms. Given the plasma parameters ($n_e \approx 5\times 10^{19}\hspace{1mm}\mathrm{[m^{-3}]}$ and $T_i \approx 1\hspace{1mm} \mathrm{[keV]}$), coulomb collisional relaxation becomes important for time scales greater  0.3 ms; hence, evolving the distribution beyond the present case requires the addition of Coulomb collisions. For lower plasma temperatures, the effect of Coulomb collisions becomes increasingly important. An important effect of collisions is the relaxation of anisotropies that arise in the presence of non-uniform magnetic fields and/or plasma heating. For example, an important physical process of interest in devices such as the linear divertor simulator MPEX is collisional relaxation and transport of fast particles created via RF plasma heating. Modelling this process provides means to determine the evolution of the distribution function in the PMI region and aid in the interpretation of exposure experiments.

The planned strategy to incorporate Coulomb collisions is based on the Monte-Carlo Fokker-Planck (FP) collision operators from references \cite{beidler_spatially_2020,caneses_effect_2020,boozer_monte_1998}. On each time step of the PIC calculation, the moments of the distribution function are calculated by extrapolating the particle states to the mesh in order to obtain spatially-resolved plasma density, temperature and plasma flow profiles. These profiles are used as inputs to the Monte-Carlo FP collision operator which takes each computational particle and simulates a collision with the background plasma based on the moments of the distribution function at the current time step. The collisions are applied in the reference frame moving with the plasma by incrementing the particle energy, pitch angle and gyro-phase according to the statistical rules determined by the FP operator. The implementation of the FP collision operator and associated physics investigation will be the subject of another paper.

\section*{Acknowledgement}
This work supported by the US. D.O.E. contract DE-AC05- 00OR22725 and sponsored by the Laboratory Directed Research and Development Program of Oak Ridge National Laboratory, managed by UT-Battelle, LLC, for the U. S. Department of Energy. The authors would also like to acknowledge  Dr. L. Carbajal Gomez, UNAM, Mexico for providing support and various physics discussions during the development of new aspects of this code.
\section*{References}
\bibliographystyle{ieeetr}  
\bibliography{mirror}
\appendix

\section{Solution to the mirror equilibrium in the paraxial limit}
\label{EqProfiles}

\subsection{Paraxial mirror equilibrium}
\label{AppA}
The paraxial mirror force balance equation is described in section \ref{GBal} (Particle and momentum transport) and for clarity is presented below in Eq. \ref{ParForceBalance}. Moreover, as described in reference \cite{Cohen1986}, the perpendicular pressure equilibrium is given by Eq. \ref{PerForceBalance}, where $B_0$ is the externally applied magnetic field in vacuum and $B$ is the beta-corrected magnetic field due to the finite perpendicular plasma pressure.
\begin{equation}
 \frac{\partial P_\parallel}{\partial x} + \frac{P_\perp - P_\parallel}{B} \frac{\partial B}{\partial x}=0
 \label{ParForceBalance}
\end{equation}\begin{equation}
 \frac{B^2}{2\mu_0} = \frac{{B_0}^2}{2\mu_0} - P_\perp
 \label{PerForceBalance}
\end{equation}Moreover, following the framework presented in reference \cite{Cohen1986}, two macroscopic stability conditions are imposed to ensure well posedness of the equilibrium problem. Let $Q$ be a quantity defined by Eq. \ref{Q}. Stability with respect to the fire-hose mode (lengthwise buckling of the magnetic field lines) requires $Q>0$ and with respect to the mirror mode requires Eq. \ref{stability2}.
\begin{eqnarray}
Q = \frac{B^2}{2\mu_0}+P_\perp + P_\parallel
\label{Q}\\
B^2 \frac{\partial}{\partial B} \bigg( \frac{Q}{B} \bigg)>0
\label{stability2}
\end{eqnarray}

\subsection{Solution to the paraxial equilibrium}
A special class of solution to Eq.(\ref{ParForceBalance}) and Eq.(\ref{PerForceBalance}) can be obtained by describing the parallel pressure $P_\parallel$ as a function of  magnetic field $B$ as described in reference \cite{Cohen1986}. Based on this, Eq. \ref{ParForceBalance} can be re-written as in Eq. \ref{ParForceBalance_B}
\begin{equation}
P_\perp  = -B^2 \frac{\partial}{\partial B} \bigg( \frac{P_\parallel}{B} \bigg)
\label{ParForceBalance_B}
\end{equation}Following the development in reference \cite{Cohen1986}, the parallel pressure $P_\parallel(B)$ is described as a $4^{th}$ degree polynomial of the magnetic field $B$ as shown in Eq. \ref{Ppar_B}. Inserting Eq. \ref{Ppar_B} into Eq. \ref{ParForceBalance_B}, leads to an expression for the perpendicular pressure $P_\perp (B)$ as shown in Eq. \label{Ppar_B}.
\begin{eqnarray}
P_\parallel (B) = a +bB+cB^2 + dB^4
\label{Ppar_B}\\
P_\perp(B) = a-cB^2 -3dB^4
\label{Pperp_B}
\end{eqnarray}Substituting the perpendicular pressure $P_\perp(B)$ into the perpendicular pressure balance Eq. \ref{PerForceBalance} leads to the bi-quadratic equation for the beta-corrected magnetic field as shown in Eq. \ref{betaB}, where the coefficients $a, c, d$ are the same terms appearing in Eq. \ref{Ppar_B} and \ref{Pperp_B}. This bi-quadratic describes the beta-corrected magnetic field $B$ in the presence of finite plasma pressure.
\begin{equation}
B^4(3d) + B^2 (c-\frac{1}{2\mu_0}) + (\frac{B_0^2}{2\mu_0} - a) = 0
\label{betaB}
\end{equation}To solve for the profiles $P_{\parallel}$  and $P_\perp$  which satisfy the paraxial mirror equilibrium Eq. \ref{ParForceBalance}, the coefficients $a,b,c$ and $d$ in Eq.(\ref{Ppar_B}) need to be determined subject to appropriate boundary conditions. The boundary conditions to constrain the coefficients $a,b,c$ and $d$ are the plasma pressures (parallel and perpendicular) at the (1) central cell and (2) mirror throats. This produces a total of four boundary conditions. Using these boundary conditions, the beta-corrected magnetic field at the central cell and mirror throats are calculated using Eq. \ref{betaB_boundary}, where $i=1,2$ corresponds to central and mirror throat regions. 
\begin{equation}
\frac{B_i^2}{2 \mu_0} = \frac{B_{0i}^2}{2\mu_0} -p_{\perp i}
\label{betaB_boundary}
\end{equation}The four boundary conditions lead to four equations with four unknown coefficients. This leads to the matrix equation shown in Eq. \ref{matrixEq} whose solution determines the values of the coefficients $a,b,c$ and $d$ in Eq.(\ref{Ppar_B}) and Eq. \ref{Pperp_B}. Once these coefficients have been determined, the beta-corrected magnetic field profile $B(x)$ can be calculated. Finally, using $B(x)$  the perpendicular $P_\perp(x)$  and parallel pressure $P_\parallel$(x) profiles as a function of physical space can be obtained. The resulting beta-corrected magnetic field $B(x)$ , perpendicular $P_\perp(x)$  and parallel pressure $P_\parallel$(x) profiles satisfy the paraxial mirror equilibrium equations Eq. \ref{ParForceBalance} and Eq. \ref{PerForceBalance}.
\begin{equation}
\left[\begin{array}{c}
P_{\parallel1}\\
P_{\perp1}\\ 
P_{\parallel2}\\
P_{\perp2}\\ 
\end{array}\right]
 = 
\left[\begin{array}{cccc}
1 & 0     & -B_1^2 & -3B_1^4 \\
1 & B_1 & +B_1^2 & +B_1^4\\
1 & 0     & -B_2^2 & -3B_2^4 \\
1 & B_2 & +B_2^2 & +B_2^4 \\
\end{array}\right]
\left[\begin{array}{c}
a\\
b\\ 
c\\
d\\
\end{array}\right]
\label{matrixEq}
\end{equation}

\section{Construction of a 4-D probability density function}
\label{4D-pdf}
In this section, a 4-D Probability Density Function (PDF) is constructed using the parallel and perpendicular pressure profiles ($P_\parallel$ and $P_\perp$)  computed from the paraxial mirror equilibrium calculation outlined in Appendix \ref{AppA}. For completeness, a plasma drift profile $\mathbf{U}$ is also included in the construction of the 4-D PDF.  The pressure profiles needed to construct the 4-D PDF are expressed in the form shown in Eq. \ref{InputPressure}, where $i$ is an index for the coordinate ($x,y$  or $z$), $n(x)$ is the particle density spatial profile and $T_i(x)$ is the temperature associated with the $i^{th}$ coordinate.
\begin{equation}
\label{InputPressure}
P_i(x)=n(x)T_i(x)
\end{equation}The PDF is constructed using a product of normal distributions for each degree of freedom in velocity space as given Eq. \ref{ND_1D}, where $i$  is an index for the coordinate ($x,y$ or $z$),  $T_i$ is the effective temperature along the $i^{th}$ coordinate, $U_i$ is the bulk plasma flow. The normal distributions satisfy the condition shown in Eq. \ref{ND_ID_prob}. 

\begin{equation}
\label{ND_1D}
f_i(v_i) = \bigg( \frac{m_\alpha}{2\pi T_i} \bigg)^{1/2} \exp \bigg( \frac{-m_\alpha(v_i-U_i)^2}{2T_i} \bigg)
\end{equation}\begin{equation}
\label{ND_ID_prob}
\int_{-\infty}^ {+\infty } f_i(v_i) d^3v =1
\end{equation}Using the normal distributions in Eq. \ref{ND_1D}, the 4-D PDF is given by Eq. \ref{PDF_4D}, where $g(x)$ is given by Eq. \ref{gx} and $h(x,\mathbf{v})$ by Eq. where $i$  is an index for the coordinate ($x,y$ or $z$). In Eq. \ref{gx}, $n_\alpha(x)$ is the spatial profile of the number density for the species $\alpha$ and $N_R^\alpha$ is the total number of real particles for species $\alpha$.
\begin{equation}
\label{PDF_4D}
f(x,\mathbf{u}) = g(x)h(x,\mathbf{v})
\end{equation}\begin{equation}
\label{gx}
g(x) = \frac{n_\alpha(x)}{N^\alpha _R}
\end{equation}\begin{equation}
\label{hxv}
h(x,\mathbf{v}) =\prod_i f_i(x,v_i)
\end{equation}The moments of the function $h(x,\mathbf{v})$  are shown in Eq. \ref{int_hvx} to Eq. \ref{int_hxv_w}, where $\mathbf{u}$ is the bulk plasma flow vector and $\mathbb{T}$ is the temperature tensor given in Eq. \ref{T_tensor}. 
\begin{equation}
\label{int_hvx}
\int_{-\infty}^{+\infty}h(x,\mathbf{v}) d^3v=1
\end{equation}\begin{equation}
\label{int_hxv_v}
\int_{-\infty}^{+\infty} h(x,\mathbf{v})\mathbf{v}d^3v=\mathbf{u}
\end{equation}
\begin{equation}
\label{int_hxv_w}
\int_{-\infty}^{+\infty} h(x,\mathbf{v})\mathbf{(v-u)(v-u)}d^3v=\mathbb{T}/m_\alpha
\end{equation}
\begin{equation}
\label{T_tensor}
 \mathbb{T}=
  \left[ {\begin{array}{ccc}
   T_x & 0 & 0 \\
   0 & T_y & 0 \\
   0 & 0 & T_y\\
  \end{array} } \right]
\end{equation}Using Eq. \ref{int_hvx} to Eq. \ref{int_hxv_w}, it can be shown that the moments of the 4-D PDF (Eq. \ref{PDF_4D}) return the pressure profiles used as inputs (Eq. \ref{InputPressure}) to the PDF. In other words, the 4-D PDF in Eq. \ref{PDF_4D} satisfies the paraxial mirror equilibrium equation presented in Eq. \ref{ParForceBalance}. This PDF is used to initialize the distribution of particle states $(x_i,\mathbf{v}_i)$  in the hybrid PIC simulation at $t=0$ using a Metropolis-Hasting algorithm\cite{Metropolis1953, Hasting1970,Calderhead2014}.

\end{document}